# CAN LINKING THE RECALL SYSTEM TO ADDICTION ENABLE A BETTER UNDERSTANDING OF THE DOPAMINERGIC PATHWAY?

Dozie Iwuh
**Augustinian Institute Makurdi, Benue State, Nigeria.**
+2348030635406
Registered501@gmail.com

**Abstract**

*Human addiction, as a learned behaviour, has and is constantly being treated psychologically, with specific and timely interventions from Neuroscience. We endorse that human addiction can receive further boost as regards treatment, when we firmly understand how it works from a quantum scale. This is majorly because the dopaminergic pathway (DP) that is well elaborated in the brain of every addict is connected to the memory pathway. This further implies that the recall process in the brain of the addict, as regards his/her addiction is fully functional in line with the pleasure that arises from the element of his/her addiction. This dopamine-led pathway shows itself as prominent in what pertains to addiction, this is because of the role it plays in reward. As a neurotransmitter, dopamine flickers when reward is in the offing. It should be noted that a full understanding of the dimensions of addiction in the human person has not be attained to, therefore, we seek to add to this ongoing research, by considering excerpts arising from Quantum Field Theory. We are introducing excerpts from QFT, because DP, is an attendant element in the process of reward and motivation. In clear terms, we are alluding that it all begins with the memory.*

**Human Addiction: An Explanation**
According to T.M. Powledge, "Addiction is… seen to be a brain disease triggered by frequent use of drugs that change the biochemistry and anatomy of neurons and alter the way they work" (Powledge, 1999) The National Institute on Drug Abuse maintains that addiction (in reference to drugs),

> is a chronic disease characterized by drug seeking and use that is compulsive, or difficult to control, despite harmful consequences. (While) the initial decision to take drugs is voluntary for most people… repeated drug use can lead to brain changes that challenge an addicted person's self-control and interfere with their ability to resist intense urges to take drugs. These brain changes can be persistent, which is why drug addiction is considered a "relapsing" disease – people in recovery from drug use disorders are at increased risk of returning to drug use even after years of not taking the drug (NIDA, 2018).

The descriptive definition above, provided by the NIDA, limits addiction to drug abuse. Yet this paper insists that addiction extends beyond the abuse of drugs, and it covers a wide range of harmful practices arising from the brain's release of dopamine that facilitates the euphoric feeling of a reward to the human being. That is to say that addictive elements will include (to add to the abuse of drug), gambling, pornography, drinking, to name a few. The release of dopamine is referred in explanatory terms by scientists to as the reward pathway in the brain. Scientist, according to Powledge, believe that it is in the activation of this reward pathway which spurs motivation into the incentive to learning and repeating an adaptive behaviour, that is called reinforcement (Powledge, 1999). The brain has several distinct dopamine pathways, one of which plays a major role in the motivational component of the motivated reward behaviour; therefore the anticipation of a certain kind of reward increases the level of dopamine in the brain (Berridge, 2007). What is being said here is that every addiction has a neural pathway, from whence the dopamine hormones uses as a means of operation. A well studied element of addiction is in drug abuse (now although this paper does not focus solely on drug addiction, it is necessary to note this point). Drug abuse, also known as the abuse of psychoactive substances witnesses to behavioural patterns which is indicated by drug tolerance, withdrawal symptoms, impairment, distress and the continued take of the drugs without regards to the negative consequences (Willuhun et al., 2010).

Empirically speaking, drug abuse affects dopamine neurotransmission. According to this evidence, people that never abused any drugs, showed an enhanced dopamine levels in the striatum, when they were exposed to these addictive substances such as cocaine and amphetamine (Volkow et al., 1997a; Drevets et al., 2001). However, it was noticed that a decreased striatal dopamine responses were reported in drug abusers (Volkow et al., 1997b). This was also true for individuals with a history of abuse of alcohol (Volkow et al., 1996), cocaine (Volkow et al., 1990), heroin (Wang et al., 1997), or methamphetamine (Volkow et al., 2001), as they showed lower levels of dopamine receptor binding when compared to nonabusers (Volkow et al. ,2004). Together, these imaging studies suggest an involvement of dopamine neurotransmission in the acute and long-term effects of abused drugs (Willuhun et al., 2010). While other addictive habits and trends are given less focus, maybe owing to the fact that their damaging effects are not as bad as that of drug abuse, it should be noted that the neural dynamics as displayed by the drug abuse addiction, is also seen in that of every other addiction that is not related to drug.

**The Dopaminergic Pathway**

Dopamine are a specific type of neurotransmitters that are involved in the reward and gratifying feel that the human person experiences. These neurotransmitters are located in the axonal part of the nerve cells and they flicker, thereby encouraging the push towards experiencing a reward-ful experience. This power of this neurotransmitter is witnessed in learning and motivation(Wise-Jordan, 2021). That is to say that that which intensifies the "good-feel" in the individual is aligned to more, but that which diminishes this, is avoided. We should thus note that any human activity that is gratifying to the individual, will be adhered to; it will be learned, then the memory of reward, will enable more as regards seeking to ensure that the reward is always felt. It is in the recalling of a memory of a reward that this DP is initiated and transmitted to that part of the human brain, which occasions action initiation. There are four DP, they

include:
1. Mesolimbic: This pathway begins at the ventral tegmental area VTA, and projects to the nucleus accumbens. It is in the nucleus accumbens that the dopamine mediates the feelings of gratification and reward. An over stimulation of the nucleus accumbens, will lead to addiction, to that which one derives pleasure from.
2. Mesocortical: This originates in the VTA, and is then transmitted to the prefrontal cortex, which is involved in cognition, memory and decision making.
3. Nigrostriatal: This starts in the substantia nigra and goes to the caudate, putamen and parts of the basal ganglia. It is connected to motor planning.
4. Tuberoinfundibular: This begins in the arcuate and periventricular nuclei of the hypothalamus and projects to the infundibular region of the hypothalamus, in particular the median eminence. This dopamine functions in inhibiting prolactin release.

From the aforementioned, it is clear that the dopamine neurons DA has a well defined construct in the brain, as contained in the DP.

**The Recall Process and the Dopaminergic Pathway**

It is to be noted that drug-induced enhancements in dopamine overflow in the nucleus accumbens, that is to say the dopaminergic pathway experiences a rush, when it is enhanced by Drugs (Di Chiara and Imperato, 1988), which has yielded results of hyperactivity that is drug induced, more to this, the sensitization of hyperactivity has been found to be associated with an increased capacity of drugs to enhance mesoaccumbens dopaminergic activity (Kalivas and Stewart, 1991; Pierce and Kalivas, 1997; Vanderschuren and Kalivas, 2000). With all these fitted together, we now come to know that repeated exposure to drugs, as in its usage, causes the brain motivational pathway to become persistently hyper-responsive and overtly sensitized to drugs and its stimulating effects. The sensitization of the mesoaccumbens dopamine system, leads to an excessive desire of the drugs and its associated good feel stimuli (Louk and Pierce 2010). Although still unclear, it is thought that the Dopamine system (DP), participates in the forming and expression of memory (Lin et al. 2020).

The anatomy of the DP, which contain dopamine neurons (DA), in the PreFrontal Cortex PFC, is quite similar in all mammals and birds (Durstewitz et al., 1998; Bjorklund and Dunnett, 2007). We have already noted that DA neurons, originate in several neighbouring midbrain nuclei, including the substantia nigra pars compacta (SNc; A9) and the ventral tegmental area (VTA; A10), these are the ones that projecct to the PFC (Puig, Rose, Schmidt and Freund, 2014). The PFC, has been thought to play an integral role in the encoding, updating and maintaining of internal representations of elements in the working memory. More to this, the PFC has also been noted to play an all important role in the dynamics of the human memory. Therefore the DP moves from its places of origin to the PFC, where it elicits the necessary action as regards the hyperactivity in the individual.

**Does The Recall Process Initiate the DP?**

Even though work in this area is yet unfolding, this research proposes that there is a DP and this comes alive after a memory is recalled. A fitting way to understanding the recall process is provided for in quantum field theory (QFT).

Before we delve into this, we need to note that we do endorse the nonlocality of brain functioning, as first earmarked by K. Lashley in the forties, who in the words of Pribram notes that "all behaviour seems to be determined by masses of excitation, by the form or relations or proportions of excitation within general fields of activity, without regard to particular nerve cells (Vitiello and Alfinito, 2000; Pribram, 1991). In the model of the brain according to QFT, the recording of a memory is obtained by the condensation of the dipole wave quanta (dwq) arising from the in-coming information that breaks the symmetry, leading to a phase shift. The brain being a dissipative system encounters memory states that have long-finite life times. As in every dissipative system, the brain should have a canonical quantization that requires the doubling of the degrees of freedom of the system.

Let $A_k$ and $\tilde{A}_k$ denote the dwq mode and its doubled mode respectively. The suffix $k$ is the momentum. The $\tilde{A}$ mode is the time-reversed image, of the A mode and it represents the thermal bath. The $A_k$ and $\tilde{A}_k$ are associated with the damped oscillator modes and their time reversed image respectively. The reason why the damped oscillator modes is used is majorly because it is a dissipative system that has the same properties as the brain system, and owing to the fact that the human brain has not undergone any validly living experiment, we use homologues when it pertains to the brain.

According to the damped oscillator modes, the couple of the classical equations to be quantized is

$$\ddot{u} + L\dot{u} + w^2 u = 0,$$
$$\ddot{i} - L\dot{i} + w^2 i = 0, \qquad (1)$$

Where u and i variables are related to the $A_k$ and $\tilde{A}_k$. When we consider the memory, dissipativity requires that such memory state, denoted by the vacuum $|0>_N$ is a condensate of equal number of $A_k$ and $\tilde{A}_k$ modes, for any $k$: this requirement ensures that the energy exchange between the system and its bath is balanced. Therefore the difference between the number of tilde and non-tilde modes has to be zero, that is to say that $NA_k$ and $N\tilde{A}_k = 0$, for and $k$. The label $N$ in the vacuum symbol $|0>_N$ specifies the set of integers $\{NA_k$, for any $k\}$ and this defines the initial value of the condensate, which is the code number associated to the information recorded at $t = 0$. The brain's ground state, that is the vacuum state can be represented as a collection (or superposition) of the full set of memory states $|0>_N$ for all $N$ (Vitiello and Alfinito, 2000). The recall process is thus considered to be the excitation of the dwq modes under the incoming stimulus, which is in fact the replication of the original signal that created the memory initially. Whenever the dwq modes are in this state of excitation, consciousness is attained to and the replication signal acts as a probe to which the brain "deciphers" the printed information (Vitiello and Alfinito, 2000) (the printing here is yet known as the recording of the memory). The replication signal arising from the bath is represented in terms of $\tilde{A}$-modes, these act to read the address of the information to be recalled. In sum we can state that

> Recall of a memory is thought in QBD to take place as follows. A new input comes in which is a "replication signal" of the signal that originally provoked the memory trace in the form of Goldstone symmetrons in vacuum states. The replication signal can be the same signal as the original one, or part of it, or even a signal with which the original one has been previously superposed. The case of superposition provides association. The interaction between the replication signal and the symmetron memory trace restores the original corticon dynamics, thereby recovering the memory (Ricciardi and Umezawa, 2004).

Memory (with particular reference to semantic and procedural memory) is enhanced more by rehearsal and recall (on the contrary, episodic memory introduces errors sometimes when the recall is made, and this explains the phenomenon of false memories) and more to that when it is attached to a particular "good feel" emotional state. There is empirical evidence to the fact that experiences that are emotionally arousing are better memorized (Brown and Kulik, 1977; Conway, 1995; McGaugh, 2003; Reisberg and Hertel, 2003). More to this, experiences that are dark and gloomy during the point of their recording, are well remembered that those that occur on routine days (Conway, 1995); this is also applicable to those that are recorded with a good feel to it. The aforementioned is further noted by Francis Bacon who stated that "memory is assisted by anything that makes an impression in a powerful passion, inspiring fear, for example or wonder, shame or joy (Bacon, 2000). According to Hebb, who proposed a dual-trace hypothesis of memory formation, memories are initially based on the reverberation of neural circuits, and long-term memory results from synaptic changes induced by the neural reverberation (Hebb, 1949); this further notes that memories are formed better after an experience. A definitive printing or recording of the memory is

> understudied by a physical process that fundamentally and totally involves the phase transition from disordered dynamics to ordered one which firmly refers to that of the corticons. And the memory recalling is maintained by a physical process of the symmetry being restored to the ordered dynamics, following the creation process of the Goldstone mode or Goldstone boson (Jibu and Yasue, 1995).

All these empirical analyses emphasize the point that before any memory is called back, there should necessary be an in flow of information that is similar to that which initially contributed to the recording of the memory. Yet it should be noted that emotions attached to memory, makes the printing process a longer lasting one.

**CONCLUSION: A Better Understanding of the DP?**

We assert here that the classical manifestation of the dopamine pathway arises when there is a phase transition that is initiated by a break in symmetry spontaneously. The DP is firmly tied to the memory recall process. According to G. Vitiello, the memory recall process, does not lie with the neuron or any other macro-scaled cell or unit (as these cannot be considered as quantum, even if they are microscopic). Rather what they put into consideration were the corticons, which are dynamical variables that are "able to describe stationary or quasi-stationary states of the brain" (Vitiello, 1995). The breaking of symmetry is what creates and initiates the dopaminergic neural pathway in the addict. It should be noted that initially, this neural pathway is nonexistent (the addict to his addiction). But when it is created, it only requires further activity on that line, therefore strengthening the neural line, leading to an ease of access in what pertains to the addiction.

The DP pathway in which the DA neurons flows in, is already been projected to providing an explanation to addiction, yet we allude that this is yet an area that needs more research done for the sake of more understanding to the phenomena of addiction. As it stands, there is no medical cure to this issue, the cure that is recommended is more psychological than clinical. Nevertheless, what if this can be clinically attended to? What if a deeper meaning into the quantum relevance of what occurs as regards Addiction is better appreciated? A lot of analyses has gone into addiction and its concerns with little consideration given to how memory mediates this. Therefore, we could further assert that if we fully understand the dynamics of how memory wades into DP and addiction, then maybe a clinical cure can be attempted. But until then, the research continues.

**Work Cited**